# A brief TOGAF description using SEMAT Essence Kernel

**(Working Paper, March 2019)**

David C. Múnera, Fernán A. Villa G.

Universidad Nacional de Colombia

**Abstract**— This work aims to explore the possibility of describing the enterprise architecture framework TOGAF using the Essence kernel SEMAT, see if they fit together, and if such marriage brings into lights any weaknesses of the models.

**Index Terms**—TOGAF, SEMAT, Description, Enterprise Architecture, Essence

—————— ◆ ——————

## 1 INTRODUCTION

This work aims to give an overview of TOGAF and SEMAT and raise the question of how well fitted are the two of them.

## 2 A BRIEF OVERVIEW TO THE FUNDAMENTALS OF SEMAT KERNEL

### 2.1 what is SEMAT?

SEMAT stands for Software Engineering Method and Theory, is an attempt to solve some of the problems that had plagued the industry for years in the complex endeavor that is software engineering itself. [1]

Problems like the oversensitivity of the industry to fashion instead of good practices, the lack of sound and standardized theoretical basis, the vast number of methods to choose from, the lack of experimental data and the huge gap between the academy and the industry. [1]

According to Jacobson et al a software engineering kernel is: "*A lightweight set of concepts and definitions that captures the essence of effective, scalable software engineering in a practice-independent way. The kernel forms a common ground for describing and conducting software development*" [1] an over-view of its main components can be seen at figure 1.

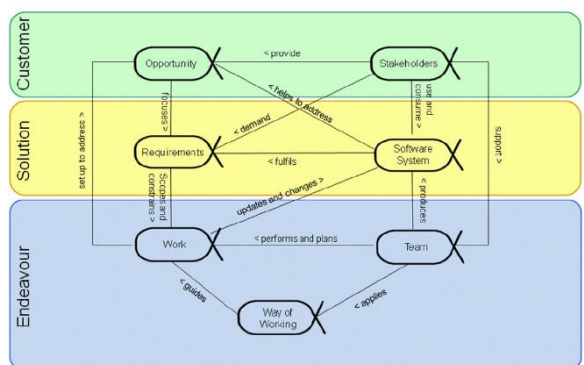

### 2.2 The kernel

#### 2.2.1 Alphas

The alphas are the essential things to work with, they provide descriptions of the kind of things a team will manage, produce and use in the process of developing, maintaining and supporting software, they are relevant to assess progress and health of the endeavor. Assessment is achieved by the monitoring of the checklist of alpha states. [1] [2]

#### 2.2.2 Activity Spaces

Representation of the kind of things to do, they provide description of the challenges a team faces during the endeavor and what the team will do to meet them. [1] [2]

#### 2.2.3 competences

Representation of the key capabilities required to carry out the work of software engineering [1] [2]

#### 2.2.4 Method

Is defined as a set of practices and a dynamic description of what has been done. [2]

#### 2.2.5 Practice

It is defined a systematic way of doing work that has a clear goal and it can be repeated. [2]

#### 2.2.6 Essence Kernel

A set of the crucial elements of software engineering, those that are key to any software engineering method [2]

### 2.3 Areas of concern

The kernel has three areas of concern, each with respect to different aspects of the engineering process, as seen in figure 1, each area of concern is related to a set of main alphas that describe what to work with in each area. The areas a represented each by a different color.

---

- *David Chaverra M Author is with the National University of Colombia, Medellin, zip 050001. E-mail: dachaverramu@unal.edu.co.*

### 2.3.1 Customer (Green)

This area is all about the use, and value of the system to be produced [1]. Here in this area the team must understand both, business and technical aspects of the domain, so they can share theirs view with the stakeholders [2]

### 2.3.2 Solution (Yellow)

This area contains the specifications and the development of the system [1]. The goal here is to capture the requirements and build a system that fulfills them [2]

### 2.3.3 Endeavor (Blue)

This area is about the team and how they go about their work. [1] How they organize themselves and manage work with the competencies available to the team [2]

## 2.4 Competencies

For each area of the software engineering endeavor, there is a set of competencies that are necessary for the work that each area requires. As seen in figure 2.

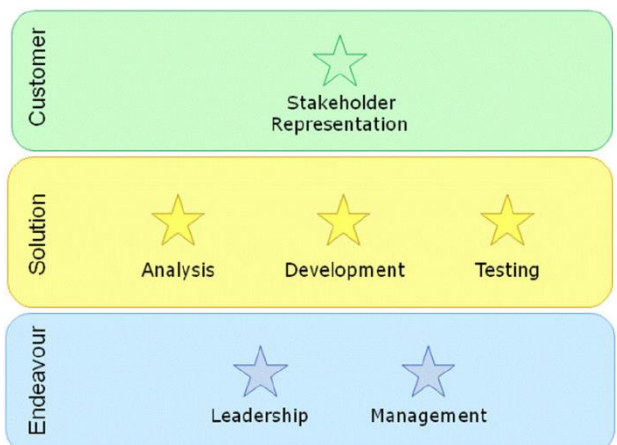

### 2.4.1 Stakeholder representation

The ability to gather, communicate and balance the needs of the stakeholders as well as represent their views. [2]

### 2.4.2 Analysis

The ability to understand the opportunities, how they relate to the needs of the different stakeholders to process them into a validated set of requirements. [2]

### 2.4.3 Development

The ability to turn the requirements into systems up to standards and norms previously agreed on. [2]

### 2.4.4 Testing

The ability to check whether the systems meets or not the requirements and specifications. [2]

### 2.4.5 leadership

The ability to move a group of people to meet the goals of their work. [2]

### 2.4.6 Management

The ability to coordinate, plan and track the work of the team, with the goal of reviewing and addressing problems [2]

## 2.5 SEMAT, Strengths and Weaknesses

### 2.5.1 Strengths

- It is a common language thus enhancing clarity and portability of the practices.
- Is extendable, allowing for the customizations needed to meet the needs of any method, or enterprise.
- Is a theoretical basis, which means it aims to be the base upon something is built, thus providing foundational knowledge to ensure any method is on track with the best practices.

### 2.5.2 weaknesses

- SEMAT is not complete, is an attempt, and a good start but is not as solid or formal as it aims to be.
- There is no clear rules and delimiters for all the SEMAT modeling components, which creates ambiguity, and make modeling too much of a perspective instead of it being fully deduced by theory.
- There is no clear way to determine which area of the effort a practice or method belongs when there is overlap, and or a way to model that overlap.
- There are elements of the kernel that could be redundant, like tasks and activities, whose definitions often overlap, and there is no clear distinction to which is the most appropriate and in which cases.

## 2.6 Why SEMAT?

The reason SEMAT was chosen for this work, is because as of now, SEMAT represents the baseline of software engineering, by being a collection of the most important concepts in the discipline, thus by making these concepts into a kernel, that is available and accepted one of the most important barriers for working in any discipline is broken, the lack of common language.

SEMAT then aims to be a set common and accepted terminology a theory, that can be used a meta model for designing practices and methods without having to start from scratch.

SEMAT is then the best tool available to verify whether a practice or a method contains all the core concepts of software engineering and make an easy to understand map of such methods.

## 3. A BRIEF OVERVIEW OF THE TOGAF ENTERPRISE ARCHITECTURE

### 3.1 TOGAF

The actual version of the TOGAF standard by the time of this work is the 9.2.

According to the Cambridge Dictionary [3]: A framework is "a system of rules, ideas, or beliefs that is used to plan or decide something" or "a supporting structure around which something can be built".

The goal of an enterprise is to deliver either goods or services to customers, as means to this end the enterprise defines processes and activities, an enterprise architecture aims to enable information technologies (IT) to integrate themselves into the after mentioned processes and activities at the level of design. [4]

TOGAF is an enterprise architecture framework engineered by the open group with focus on aligning IT goals, with business goals, coordinate efforts and define requirements.

According to the open group the main benefits of an enterprise architecture are the following: [5]
- More effective and efficient business operations
- More effective and efficient IT operations
- Better return on existing investment
- Faster, simpler and cheaper procurement

To summarize the TOGAF standard proposes a set of practices encompassed within its own methodology, that is described with its own set of concepts that server as common language and make use of some tools it describes, to produce some artifacts that helps structure the Enterprise towards a gaining something.

### 3.2 Fundamental aspects of TOGAF

The TOGAF architecture proposes a set of steps or practices that in an iterative cyclical manner according to The Open Group [5] ADM *"provides a tested and repeatable process for developing architectures"* these steps are contained if the following phases:

#### 3.2.1 The Preliminary Phase
This is an exploratory phase, that aims to understand what the organization already knows, what is it environment and how that information is going to shape the desired architecture. [5]

#### 3.2.2 Phase A. Architecture Vision
This phase is all about the scope and boundaries of the architecture and its endorsement by cooperate management, while making certain that definitions and knowledge is current and updated. [5]

#### 3.2.3 Phase B. Business Architecture
This phase describes the strategies and the environment of the architecture. This phase aims to document and update already existing elements of the business strategy while re-using existing material as much as possible and filling the gaps left by other phases. [5]

#### 3.2.4 Phase C. Information Systems Architectures
This phase aims to integrate technology and data into the architecture, address the management of said data to make that data work seamlessly with architecture. This phase also concerns itself organizing and making available the resources found in the Architecture repository. [5]

#### 3.2.5 Phase D. Technology Architecture
This phase goal is to review the already existing IT resources, reference models and technology models that will be useful for the logical al physical implementation of the architecture, also this phase aims to update the already existing architecture roadmap based on the gaps between the baseline and target architecture. [5]

#### 3.2.6 Phase E. Opportunities & Solutions
Here the goal is to sketch the architecture roadmap, based on the information gathered in previous phases, check if an incremental approach is going to be needed and, in such case, identify the architecture that will deliver it. This phase also aims to design the implementation and migration plan, which is a schedule of the actions to be undertaken to build the target architecture [5]

#### 3.2.7 Phase F. Migration Planning
The aim of this phase is to complete and refine the work done in the previous phase, verify that the work products built are coordinated with the enterprise current environment and goals and make certain that the cost of the work is understood by stakeholders. [5]

#### 3.2.8 Phase G. Implementation Governance
This phase aims to verify that the actual work implementation is in line with the architecture that has been plan and design up to this point and implement any request for change that is deemed needed. [5]

#### 3.2.9 Phase H. Architecture Change Management
This phase has the goal of varying weather the architecture built, the target architecture meets the requirements, and ensure that the cyclic nature of this phases is met by restarting the phase A [5]

#### Requirements Management
This is an unordered phase, hence the lack of letter, this phase is always being executed, it captures all requirements found and compiled during any phase to ensure that they are always up-dated and available is phases that are needed. [5]

TOGAF uses something called Architecture Repository which it uses to store different outputs generated during each of its phases, these outputs are called artifacts, these artifacts are categorized as [5]:
- Catalogs
- Matrices
- Diagrams
- Other deliverables

### 3.3 TOGAF, Strengths and Weaknesses

#### 3.3.1 Strengths
- Freely available for implementation
- Very well documented, with easy access to the documentation right on the open group webpage.
- Widely spread with 32% usage in the public sector [4] and about 50% of the private sector according to their own numbers [5].
- Huge amount of online resources, and certified training facilities.

#### 3.3.2 Weaknesses
- Just as its predecessor TAFIM who went out of market due to its projects being too big and impractical [6], TOGAF seems to follow on its steps, since the whole TOGAF standard is huge and its complete implementation is mostly impractical on many levels of enterprises.
- TOGAF has remained mostly static since its development and very little empirical data is known of how the best practices of TOGAF stand up to current enterprise implementations [6].
- TOGAF has focused too much on being a set of tools for modeling architecture instead of its original focus on being a collection of best practices, since it is still unknown where those best practices came from [6].

### 3.4 Why TOGAF?
TOGAF is broadly documented, widely spread, and most of the time taken as de-facto of how an enterprise architecture should be implemented, one of TOGAF goals is to provide a common ground of concepts that would make implementing and reusing enterprise architectures models simple and efficient, although that last statement is debatable.

The fact that TOGAF is so widely spread is a testament of fulfillment of the goal of providing common ground.
TOGAF also aimed to make its framework somewhat a meta model and to this end it has a whole toolkit of modeling, according to the open group [5] TOGAF is flexible, can be implemented alongside any other architectural framework and can even be used to extend them. TOGAF then shares similarities with SEMAT, that makes the pairing highly compatible.

## 4. REPRESENTING TOGAF USING SEMAT ESSENCE KERNEL

### 4.1 The model construct
Togaf is a cyclical reiteration of several phases, being cyclical implies that each phase must be repeated also they have a list of inputs, and outputs, and a checklist of goals that must be achieved and they are composed by a set of steps, that must be done to reach the goals. Providing a list of steps is a very systematic way of organizing work to achieve the goals, which is the definition of practice in TOGAF hence in this work each phase will be modeled as a practice an example of this is found in figure 3.

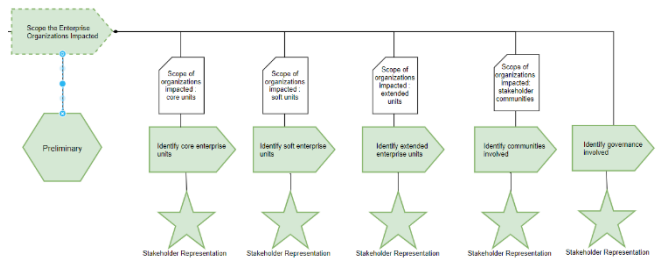

Each step within a phase encapsulates work that needs to be done, and how to get it done by a subset of activities that must be completed in order to finish the step, in order to represent this relationship this work will model each step within a phase as activity space, since activity spaces are a collection of things that have to be done, and it challenges, being each sub activity considered the how those challenges will be met.

Activities represent the actual actions that will be taken to complete work, this are modeled as activities instead of tasks because the atomic nature of tasks is not identified within TOGAF actions, since they lack any explicit definition of their coverage, and particularly problematic, they do mention which roles will be responsible, hence activity is more accurate, since it requires less detail in its definition.

For activities that are particularly large, and are explicitly broken down into smaller activities, such activities were replaced with nested activity spaces that organized their sub-activities as SEMAT activities [1] to make the model more readable, and accurate. This can be seen in figure 4

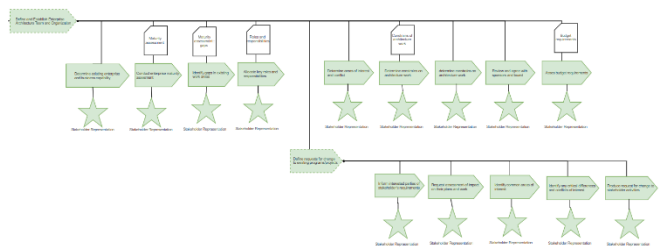

The specific output documents of each phase will be model as a work product in SEMAT, and will be anchor to its corresponding activity, if its known, with further details, if multiple activities contribute to a work product a duplicate will be added for each of them, with the specific part of the work product that the activity adds, if known, after the colons.

Competencies were added with respect to their SEMAT definitions [2], if an activity was about acquiring information, understanding stakeholder's views or processing requirements, the competency Stakeholder Representation was chosen, the analysis competency of the SEMAT kernel was reserve for those activities where the requirements processed got endorsed by stakeholders, as per its definition of "agreed" [1] requirements.

For this model, a new competency was deemed necessary since all SEMAT kernel competencies are from the perspective of a very liberal self-organized team, but that is not always the case, sometimes there are authority figures that can make decisions that constrain the way of work of the team, like authorization and application of policies. The competency Governance is defined in the scope of this work as the capability of making decisions that constrain the team's way of work. An example of the used of this competency can be found in figure 5.

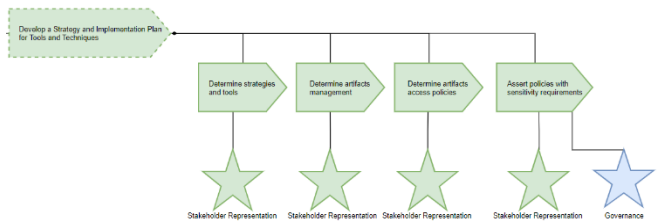

The scope of this work was mainly to answer the question of how well equipped SEMAT was to model TOGAF, and given such case, build a preliminary model of TOGAF using SEMAT, due to the sheer size of the TOGAF EAF framework, this work was constrained to only model the introductory phases of TOGAF, which are the preliminary phase, and the architecture vision phase, phase A.

### 4.2 TOGAF identified issues

Deliverables sometimes are defined in multiple places, and with different requirements, which makes hard to keep up with exactly what each artifact must contain.
TOGAF has a huge list of outputs for every phase, and an even bigger number of activities to do in each phase, problem is TOGAF most of the time does not tell you which deliverable the activity is supposed to feed.

TOGAF is very inconsistent with its documentation, especially in the steps department, that are sometime defined as a set of tasks to be done and sometimes as a nice piece of information that you must try and guess what is that step about, since neither the goals for the activity or tasks is explicit, nor there is any controlled language thus making the step highly open for personal interpretation of the information there presented.

A missed opportunity lies in the detailed definition of roles, TOGAF define the roles that are needed for a successful architecture endeavor, as well as the level and kind of competency they are required to have, but this definition is kind of wasted since is never stated which role should be responsible for what activity and, or, work product.

### 4.3 Difficulties found when modeling

SEMAT has no way of modeling the TOGAF capability framework, this framework is quite detailed, it not only tells you what capabilities a role needs to have, but also its level required [5], SEMAT on the other hand only tells you the capability itself, but has no way to mention the level of this one, which may come in handy when finetuning costs by choosing the correct capability level for the task, making sure the more capable assets focus on the correct task.

SEMAT has three areas of concern, separated by colors, by activities, alphas, and capabilities. These can be mixed during modeling; hence it would be expected to have a criterion for determining the area of concern a method would belong to, due to its mixed components, the thing is, there is no such criteria, leaving the choice purely up to the modeler, which goes against being solid.

There is no clear statement of what items it SEMAT modeling kit can be nested, but also no mention that they cannot be, this can be an issue since the modeler have to guess whether certain pairing is allowed but cannot confirm whether the model is valid.

## 5. CONCLUSIONS

SEMAT is well equipped to model the TOGAF framework even with the lack of a solid enough ability to model capabilities which TOGAF clearly having a more detailed model capacity in said area. Although not used a its full potential due to not having clear role or capability definition for neither of its activities, phases or deliverables. Apart from the architecture board, but even then, only which goals the architecture board must pursue are described.

SEMAT and TOGAF share some similarities in the ideas that gave birth to both models, like being flexible, becoming a common ground, and ensuring stakeholder satisfaction. This raises the concern as TOGAF has focused too much in the marketing of its supporting tools [6] instead of enhancing its theoretical basis, that could SEMAT be on its way towards a similar fate and if so, how can it be prevented.

It it's clear that the modeling of TOGAF using the SEMAT Essence kernel has brought into light some issues with TOGAF EAF which raises the opportunity for future work.

## 6. FUTURE WORK

The future line of this work will be to do an extensive model of the TOGAF EAF using SEMAT, create a table of the issues each of the models present, and propose solutions to said issues.